\documentclass[useAMS,usenatbib]{mn2e}
\usepackage{amsfonts}
\usepackage{amsmath}
\usepackage{amssymb,url}

\def\bea{\begin{eqnarray}}
\def\ena{\end{eqnarray}}

\usepackage[dvips]{graphicx}
\usepackage{amsmath,amssymb}
\usepackage{myaasmacros}
\usepackage{color}
\usepackage{multirow}

\title[Fermi LAT GRBs: new candidates found.]{GRB observations by Fermi LAT revisited: new candidates found.}

\author[G. I. Rubtsov, M. S. Pshirkov and P. G. Tinyakov]{G. I. Rubtsov$^{1}$\thanks{E-mail:
grisha@ms2.inr.ac.ru }, M. S. Pshirkov$^{2,3}$\thanks{E-mail:
pshirkov@ulb.ac.be} \& P.G.
Tinyakov$^{1,2}$\footnotemark[1]\thanks{E-mail:
petr.tiniakov@ulb.ac.be} \\
$^{1}$Institute for Nuclear Research of the Russian Academy of Sciences, 117312, Moscow, Russia\\
$^{2}$Universite Libre de Bruxelles, Service de Physique Theorique, CP225, 1050,  Brussels, Belgium\\
$^{3}$Pushchino Radio Astronomy Observatory, 142290 Pushchino, Russia\\
}

\begin{document}

\date{}

%\pagerange{\pageref{firstpage}--\pageref{lastpage}} \pubyear{2002}

\maketitle

\label{firstpage}

\begin{abstract}

We search the Fermi-LAT photon database for an extended gamma-ray
emission which could be associated with any of the 581 previously
detected gamma-ray bursts (GRBs) visible to the Fermi-LAT. For
this purpose we compare the number of photons with energies
$E>100~$MeV and $E>1$~GeV which arrived in the first 1500 seconds
after the burst from the same region, to the expected background.
We require that the expected number of false detections does not
exceed 0.05 for the entire search and find the high-energy
emission in 19 bursts, four of which (GRB 081009, GRB 090720B, GRB
100911 and GRB 100728A) were
previously unreported.  The first three are detected at energies
above 100~MeV, while  the last one shows a statistically
significant signal only above 1~GeV.

\end{abstract}


%%%%%%%%%%%%%%%%%%%%%%%%%%%%%%%%%%%%%%%%%%%%%%%%%%%%%%%%%%%%%%%%%

%%%%%%%%%%%%%%%%%%%%%%%%%%%%%%%%%%%%%%%%%%%%%%%%%%%%%%%%%%%%%%%%%%%
%%%%%%%%%%%%%%%%%%%%%%%%%%%%%%%%%%%%%%%%%%%%%%%%%%%%%%%%%%%%%%%%%%%

\section{Introduction}
\label{sec:intro}

Gamma-ray bursts (GRBs) are one of the brightest phenomena in the
Universe. Although the majority of GRBs were detected at energies
ranging from hundreds of keV to several MeV, they were also
observed at much higher energies up to tens of GeV
\citep{Hurley1994}. The advent of the Fermi Large Area Telescope
(LAT) with its unprecedented sensitivity
\citep{Atwood2009,Band2009} has greatly increased our
capability to study  the high-energy emission from GRBs
\citep{Omodei2009}.

The high-energy (HE) emission was detected both in the prompt and
afterglow phases of GRBs. However, its origin is still unclear: it
could be produced in the internal/external shocks via leptonic or
hadronic mechanisms, or in the process of dissipation of the Poynting
flux (e.g.,
\cite{Meszaros1994,Waxman1997,Bahcall2000,Zhang2001,Dermer2004,Fan2008,Panaitescu2008,Zhang2009,Ghisellini2010,Kumar2010,Razzaque2010,Zhang2011, Meszaros2011}).
Detailed information on the high-energy $\gamma$-ray prompt and
afterglow emission could shed light on the onset of GRB and its
immediate interaction with surrounding interstellar medium.

GRB observations in the HE band are one of the key science topics of
Fermi-LAT. The Fermi Gamma-Ray Burst Monitor (GBM) can initiate
autonomous slew of the spacecraft to provide the best conditions for
a dedicated GRB observation. Also, knowledge of time and position
of a GRB (either provided by GBM or by other observations) makes it
feasible to search for GRBs in the LAT photon database where they
would manifest themselves as spatially and temporally compact clusters of
photons \citep{Band2009}. There has been 20 LAT detections of GRBs as of
the time of writing (Feb 2011).

Observations of several very bright bursts (GRB 080916C, GRB 090510,
GRB 090902B, GRB 090926A) containing more than 100 photons with
energies in excess of $100$~MeV allowed one to study the spectral
properties of the HE emission and even their temporal evolution
\citep{Abdo2009a,Abdo2009b,Ackermann2010,Ackermann2011}. In all these
cases the HE emission demonstrated a delay (of several seconds for
long bursts, tenth of a second for the short burst GRB 090510) with
respect to the prompt emission in the sub-MeV energy range. The HE
emission also lasted much longer. The observations indicated a
significant deviation from the so-called 'Band function'
\citep{Band1993}, namely, the presence of a hard power-law component
that dominates at high energies (e.g.,
\citep{Abdo2009b}). Observations of the short GRB 090510
\citep{Ackermann2010} were especially fruitful: first, they proved the
intrinsic similarity of the HE emission between short and long bursts;
second, the detection of a photon with energy over 30~GeV made it
possible to evaluate the Lorentz factor of the jet $\Gamma\sim 1000$;
finally, the simultaneous detection of signals in different energy
ranges of GBM and LAT allowed one to put constraints on some quantum
gravity theories which predict Lorentz invariance violation and
energy-dependent speed of light \citep{Abdo2009c,Ghirlanda2010}.

In this letter we use the Fermi-LAT data to search for a high-energy
emission related to GRBs. A similar question was addressed in a number
of recent papers. In the papers by \cite{Akerlof2010,Akerlof2011} the
matched filtering technique was implemented and three detections of HE
emission were claimed. In the paper by \cite{Nakar2011} the HE
fluence associated with several bright GBM events was
constrained.

Unlike the above studies, we looked for a HE emission that could
extend over longer time spans than were examined previously and which
therefore could have been missed. For this purpose we searched the LAT
database for HE photons in two energy bands ($E>100$~MeV and
$E>1$~GeV) and selected those which came from the directions of 581
previously detected GRBs within the time window of 1500~s either
before or after the burst. We then estimated the statistical
significance of the excess by comparing the observed number of photons
with the expected background, taking into account the penalty for the
total number of trials. The detection criterion was chosen in such a
way that the entire search would give a single false detection with
the probability of 0.05. We found 4 previously unreported GRBs showing
the post-burst HE emission, and none showing the pre-burst emission.

\section{Data}
\label{sec:data}

The LAT detector has a sensitivity to photons with energy above 30
MeV with a wide field of view (FOV) of $\sim2.4$~sr and the
effective area of up to 9500 cm$^2$. The detector angular
resolution is a function of the incident photon energy and the
'event class' which is determined by the set of reconstruction
cuts~\citep{Rando2009}. It also depends on the angle
between the instrument axis and the arrival direction of a photon,
but for the purpose of present analysis, we use the
mission-average value. An extensive review of the instrumental
capabilities can be found in \citep{Atwood2009}.  In this letter
we use the LAT weekly all-sky data that are publicly available at
Fermi mission
website\footnote{http://fermi.gsfc.nasa.gov/ssc/data/access/}. The
analysis covers the time period of 127 weeks from August 04, 2008
to January 07, 2011, corresponding to mission elapsed time (MET)
from 239557417~s to 316111862~s.

We use the 'diffuse' event class and impose an Earth relative
zenith angle cut of $105^\circ$. We discard the photons from
'transient' and 'source' classes. The inclusion of these classes
would decrease the signal-to-background ratio which becomes an
important factor for our comparatively long time window (see
\cite{Akerlof2011} for detailed discussion of event classes with
respect to GRBs). We do not require the rocking angle cut of
$52^\circ$ in order to keep photons observed in the pointing mode,
including repoint requests caused by the GRB trigger.

For the spatial selection of photons we adopt the 95\% containment
angle $\alpha_{95}(E,v)$ corresponding to the point spread
function (PSF) for the ``diffuse' class photons
\citep{Rando2009,Abdo:2009gy,Burnett:2009ig}. This angle depends
on both the photon energy and the conversion type $v$. The latter
takes two discrete values: 0 and 1 for front and back converted
photons, respectively.

In our analysis we used the times and coordinates of the GRBs detected
by other instruments such as
GBM\footnote{http://fermi.gsfc.nasa.gov/ssc/data/access/}
\citep{Meegan2009},
Swift\footnote{http://swift.gsfc.nasa.gov/docs/swift/} \citep{Swift},
INTEGRAL\footnote{http://www.isdc.unige.ch/integral/}
\citep{INTEGRAL}, MAXI \citep{MAXI} and Konus-Wind \citep{KONUS}. We
have compiled two non-overlapping list of GRBs. The first list
included 605 GRBs detected by Fermi GBM and the second 279 GRBs
detected by other instruments only.  Of these, 444 and 137 GRBs,
respectively, were in the Fermi-LAT FOV at least for some part of
1500~s after the burst. Note that although we used time stamps from
the original observations, localizations in many cases were provided
by much more precise follow-up observations; these data were obtained
through the Gamma-ray bursts Coordinates Network (GCN;
http://gcn.gsfc.nasa.gov/).

\section{Method}
\label{sec:method}

\begin{table*}
\begin{center}
\begin{tabular}{|c|c|c|c|c|c|c|c|c|c|}
\hline GRB name  & \multicolumn{4}{|c|}{$E~>~100$~MeV} &
\multicolumn{4}{|c|}{$E~>~1$~GeV}& \\
~ & $B$ & $n$ & $p$ &$\mathcal{E},~10^5~\mathrm{cm^2s}  $& $B$ & $n$ & $p$ &$\mathcal{E},~10^6~\mathrm{cm^2s} $& $t$, s \\
\hline
080916C & 3.9  & 125 &  5.7e-138&2.16 & 0.065  & 18 & 6.4e-38 &5.55&1500\\
*081009 & 1.9 & 11 & 4.8e-6  & 1.96 & 0.032  & 1  & 0.031     &5.1&1430\\
081024B & 0.32 & 5 & 2.3e-5  & 0.36  & 0.0044 & 1  & 4.4e-3 &0.74& 1500\\
090217A & 0.82 & 10 &  1.9e-8& 0.75  & 0.0090 & 1  & 9.0e-3 &2.20&600 \\
090323 & 1.2 & 31 & 2.6e-32  & 2.42 & 0.012 & 4  & 9.2e-10 & 3.58&1210\\
090328 & 3.5  & 28 & 1.4e-16  & 5.88& 0.043  & 8  & 2.6e-16 &8.10&1500\\
090510 & 2.8  & 121 & 7.9e-148 &7.20 & 0.036  & 27 & 1.3e-67 &9.43&1500\\
090626 & 1.1 & 10 & 3.1e-7    &1.15 & 0.020 & 0  & 1.0     &2.88&750\\
*090720B & 4.6 & 16 & 2.4e-5  & 0.63  & 0.070  & 0  & 1.0    &3.26&1500 \\
090902B & 2.6  & 166 & 8.2e-231& 4.70& 0.036  & 33 & 2.1e-85 &6.41&1500\\
090926A & 0.42 & 130 & 9.2e-270& 0.37  & 0.0051 & 20 & 6.7e-65 &1.49&530\\
091003A & 3.9  & 25 & 7.6e-13  & 6.99 & 0.034 & 3  & 6.1e-6 &9.16&1500\\
091031 & 2.5 & 13 & 2.2e-6& 3.96& 0.028 & 1  & 0.027 &7.31&1500\\
100116A & 2.1  & 14 & 5.8e-8& 1.54   & 0.033 & 4  & 4.8e-8&3.41& 820\\
100414A & 2.9  & 20 & 6.2e-11& 6.32  & 0.039  & 4  & 9.5e-8 &8.59&1450\\
100724B & 0.43 & 6 & 6.3e-6& 0.33& 0.0046 & 0 & 1&1.18&1500\\
*100728A & 4.2  & 10 & 0.010&  0.55 & 0.065  & 4  & 7.1e-7 &8.17&1500\\
*100911 & 0.059 & 3 & 3.3e-5&0.016 & 0.0002 & 0 & 1 &0.74&460\\
101014A & 0.92 & 8 &  5.6e-6&  0.88& 0.0048 & 0 & 1&1.47&1500\\
\hline
\end{tabular}

\end{center}
\caption{List of Fermi-LAT GRBs showing extended high-energy
emission. $B$ is  the expected background, $n$ is the observed
number of photons,  $p$ is  the probability that the signal is
fluctuation of the background, $\mathcal{E}$ is the
exposure, and $t$ is total time duration within 1500 s interval
when the angle between boresight of the telescope and the GRB
position was less than $65^{\circ}$. Previously unreported
candidates are marked  with the star. \label{candlist}}
\end{table*}

\begin{table*}
\begin{center}
\begin{tabular}{|c|c|c|c|c|c|c|c|c|c|}
\hline GRB name  & \multicolumn{4}{|c|}{$E~>~100$~MeV} &
\multicolumn{4}{|c|}{$E~>~1$~GeV}& \\
~ & $B$ & $n$ & $p$ &$\mathcal{E},~10^5~\mathrm{cm^2s}  $& $B$ & $n$ & $p$ &$\mathcal{E},~10^6~\mathrm{cm^2s} $& $t$, s \\
\hline
080825C & 2.3 & 8 & 2.8e-3 &1.59& 0.037 & 0 & 1&4.44& 1330\\
081215A & 0 & 0 & 1&0 & 0 & 0 & 1&0&0\\
100225A & 2.9 & 4 & 0.33&4.06 & 0.042 & 0 & 1&7.22&1500\\
100325A & 2.6 & 6 & 0.049&5.03& 0.026 & 0 & 1&8.07&1500\\
100707A & 0 & 0 & 1&0 & 0 & 0 & 1&0&0\\
\hline
\end{tabular}
\end{center}
\caption{List of previously detected Fermi-LAT GRBs missed by our
algorithm. \label{faillist}}
\end{table*}

The key quantity in our analysis is the probability $p$ that the
observed HE emission from the direction of a given GRB is a
fluctuation of the background. If this probability is smaller than
the certain threshold, we claim the detection of the high energy
emission from that burst. The significance threshold is obtained
by requiring that the number of false detections does not exceed
0.05 in the entire set. Taking into account that the total number
of bursts is 581 and counting two energy ranges as independent, one
obtains the following condition:
\begin{equation} \label{condition}
p~<~5\times 10^{-5}
\end{equation}
for either of the two energy regions.

The probability $p$ for a given burst and given energy threshold
$E_0$ is calculated as follows. Let $t_b$, $l_b$ and $b_b$ be the
trigger time and galactic coordinates of a GRB. First, we determine
the observed number of photons $n$ above the energy $E_0$ by
counting photons satisfying the following conditions:
\begin{align}
E &> E_0\,,\nonumber \\
\alpha(l,b,l_b,b_b) &
< \sqrt{\alpha^2_{95}(E,v)+\alpha_{\rm GRB}^2}\;,
\label{cndA}\\
t_b &\le t \le t_b + 1500~\rm{s}\,,\nonumber
\end{align}
where $t$, $l$, $b$, $E$ and $v$ stand for arrival time, coordinates,
energy and conversion type of a photon, $\alpha(l,b,l_b,b_b)$ is the
angular separation between photon and GRB, and $\alpha_{\rm GRB}$ is the
GRB pointing error. The energy threshold $E_0$ is either 100~MeV or
1~GeV. These conditions select photons with energies larger than $E_0$
that arrived within 1500~s after the burst from the region of interest
(ROI). The latter is a circle with the energy-dependent radius
determined by the two contributions: the error of the photon arrival
direction and the error of the GRB position. Usually the first
contribution dominates. The error of the GRB position was
taken to be equal to 1$^{\circ}$ in the case of GBM bursts and
0.5$^{\circ}$ in the case of bursts detected by Swift. Errors for all
other bursts were determined individually from the GCN website.
The observed pre-burst photons are selected by an obvious modification
of the conditions (\ref{cndA}).

Next, we calculate the expected background $B$ corresponding to the
energy $E>E_0$.  Since GRBs are exceptional events, for the background
calculation we may use the photons from the same spatial region for
the entire duration of the mission. Thus, the background is given by
the total number of photons observed from the ROI during the whole
mission, multiplied by the ratio of the exposure corresponding to
1500~s after the burst to the total exposure of ROI. The calculation
which takes into account the energy dependence of ROI is presented in
the Appendix \ref{app:bgnd}.

Finally, having calculated the observed number of photons $n$ and the
expected background $B$, the probability $p$ for the GRB in question
is calculated from the Poisson distribution,
\[
p = \mathcal P(B, n),
\]
where $\mathcal P(B,n)$ is the probability to observe $n$ or more
events at $B$ expected. If this probability satisfies the condition
(\ref{condition}) for at least one of the two energy regions of
interest, we have a detection and include the corresponding GRB in the
detection list,  Table~\ref{candlist}.

\section{Results and conclusions}
\label{sec:results}

Applying the method of Sect.~\ref{sec:method} to 581 GRB we have
achieved 19 detections of the post-burst HE emission, of which 4
(namely, GRB 081009, GRB 090720B, GRB 100728A and GRB 100911) were
previously unreported. All detections correspond to GRBs present in
the Fermi-GBM part of the GRB list. No pre-burst HE emission was
found.  Of the new detections, GRB 100728A demonstrated particularly
bright and long HE afterglow: 4 photons with energy $>1$ GeV were
observed vs. 0.065 expected from the background, the chance
probability being $p=7\cdot 10^{-7}$. Photon arrival times relative to
the burst trigger time are 711.3~s, 713.8~s, 1161.0~s and
1342.5~s. GRB 081009, GRB 090720B and GRB 100911 were observed with
the rocking angle less than 52$^\circ$, while during the GRB 100728A the
rocking angle exceeded 52$^\circ$ 1220~s after the burst.
The low exposure for GRB100911 is because that burst went behind the Earth shortly after the trigger and there exists a possibility of contamination from the bright Earth limb.

The total number of previously reported GRBs detected by Fermi-LAT is
20. Our algorithm failed in 5 cases; they are listed in
Table~\ref{faillist}. Two of them (GRB 081215A and GRB 100707A)
resided far from the Fermi-LAT axis at approximately 90$^{\circ}$
angular distance. Their original detections were made with the use of
the non-standard analysis technique \citep{GRB081215A,
  GRB100707A}. Our algorithm is based on the standard event
reconstruction and does not treat events outside of LAT FOV. The
remaining three GRBs are seen as excesses that do not satisfy
eq.~(\ref{condition}) (in the worst case of GRB 100225 the chance
probability is as large as 0.4).  This discrepancy can be
attributed to a wider time window used in our analysis and
somewhat different energy ranges.

In our procedure we have treated the background events as a Poissonian
process. This approach would fail in case of a moving gamma ray source
(the Sun or the Moon) crossing the region of interest just at the
moment of burst. We have explicitly checked that no such events
happened for the reported candidates. We have also assumed that the
background flux is stationary. This could lead to an erroneous
detection if some gamma-ray sources in the region of interest flared
at the moment of the GRB. This issue should be investigated
separately; one can use the LAT 1-year Point Source
Catalog~\citep{Fermi:2010ru} for this purpose. No known sources appear
within 95\% containment radius of all GeV photons attributed to GRB
100728A; the same is valid for GRB 081009. On the contrary, there are
numerous sources in the neighborhoods of GRB 090720B and GRB
100911. The question of a possible influence of variability of these
sources on the present analysis will be studied elsewhere
\citep{Rubtsov2011}. Finally, to test for a possible influence of the
magnetospheric flares\footnote{We would like to thank B.~Stern for
  pointing out this possibility.} we calculated the gamma-ray flux
during 1500~s after the burst in the ring between $15^\circ$ and
$20^\circ$ from its location, and compared it to the expected
background. No indication of magnetospheric flares coincident with the
reported new detections was found.

When this paper was already submitted, analysis from the
Fermi collaboration  appeared confirming detection of HE emission
from GRB 100728A \citep{Abdo:2011}.

\section*{Acknowledgements}
We are indebted to O.~Kalashev, A.~Neronov, K.~Postnov, B.~Stern,
I.~Tkachev and S.~Troitsky for helpful discussions. We would like to thank   anonymous referee for providing us with constructive comments that helped us improve the quality of the paper.
The work was supported in part by the RFBR grants 10-02-01406a,
11-02-01528a, 12-02-91323-SIGa~(GR), by the grants of the President of
the Russian Federation NS-5525.2010.2 (GR), MK-1632.2011.2 (GR),
MK-1582.2010.2 (MP), by the Ministry of Science and Education under
state contracts 02.740.11.0244 (GR), P2598 (GR) and 14.740.11.0890
(PT). The work of M.P and P.T. is supported in part by
the IISN project No. 4.4509.10. GR wishes to thank the hospitality of
ULB Service de Physique Theorique where this study was initiated. The
numerical part of the work was performed at the cluster of the
Theoretical Division of INR RAS. This research has made use of NASA's
Astrophysics Data System.

\appendix
\section{Background calculation}
\label{app:bgnd}
The background expected from the region of interest during 1500~s
after the burst is determined by the product of the differential flux
$f(E,l,b)$ from the ROI and the detector exposure in that direction.

The number of photons coming within the time period
$[t_1,t_2]$ is:
\begin{equation*}
N = \int dE\,dl\,db\,f(E,l,b) \mathcal E(E,l,b,t_1,t_2),
\end{equation*}
where $\mathcal E(E,l,b,t_1,t_2)$ is the exposure of the instrument in
a given direction and the time period $[t_1,t_2]$ at energy $E$, measured in units
of $\mathrm {cm^{2}\,s}$. This exposure is estimated using the
standard Fermi-LAT tools {\it gtltcube} and {\it gtexpcube}
(ScienceTools-v9r18p6-fssc-20101108).

The same number of photons can be written alternatively as
\begin{equation*}
N = \int dE\,dl\,db\,n(E,l,b,t_1,t_2),
\end{equation*}
where $n(E,l,b,t_1,t_2)$ is the angular spectral density of
photons detected over the period $[t_1,t_2]$. Comparing these two
equations one finds:
\begin{equation}
\label{nf}
n(E,l,b,t_1,t_2) = f(E,l,b) \mathcal E(E,l,b,t_1,t_2).
\end{equation}
Therefore, knowing $n(E,l,b,t_1,t_2)$ and $\mathcal E(E,l,b,t_1,t_2)$
allows one to estimate the differential flux $f(E,l,b)$. Under the
assumption of a time-independent flux, the best estimate is obtained
using the longest possible time period $[t_{\rm start}, t_{\rm end}]$.

We represent $n(E,l,b,t_{\rm start}, t_{\rm end})$ as a discrete distribution
originated from all the photons observed for the whole duration of the
mission $[t_{\rm start}, t_{\rm end}]$:
\begin{equation}\label{ndiscrete}
n(E,l,b,t_{\rm start}, t_{\rm end}) = \sum\limits_{i=1}^{N}
\delta(E-E_i)~\delta(l-l_i)~\delta(b - b_i),
\end{equation}
where $l_i, b_i, E_i$ are coordinates and energy of $i-$th photon.
For the definition of the ROI we impose energy, spatial and time cut functions
corresponding to Eq.~\ref{cndA}:
\begin{align*}
\sigma_e^{E_0}(E) &= \theta(E - E_0)\,,\\
\sigma_b(E,v,l,b) &= \theta(-\alpha(l,b,l_b,b_b) +
 \sqrt{\alpha^2_{95}(E,v)+\alpha_{\mathrm GRB}^2})\,,\\
\sigma_{t}(t) &= \theta(t-t_b)~\theta(-t + t_b+1500~\mbox{s})\,,
\end{align*}
where $\theta(x)$ stands for Heaviside step function.

Number of background photons is estimated as a sum of front
and back converted components each calculated with its own PSF and
exposure:
\begin{align}
B_{E_0} = \sum\limits_{v=0}^{1}&\int
dE\,dl\,db\,f(E,l,b)~\mathcal
E(E,l,b,t_b,t_b+1500~\rm{s}) \nonumber\\
 &\times \sigma_e^{E_0}(E)~\sigma_b(E,v,l,b)\,,
\label{bgbase}
\end{align}
where $\mathcal E^v$ stands for conversion-type dependent exposure
($\mathcal E = \mathcal E^0 + \mathcal E^1$). We substitute $f$ from
Eq.\ref{nf} and Eq.\ref{ndiscrete} to Eq.\ref{bgbase}:

\begin{align*}
B_{E_0} = \sum\limits_{v=0}^{1} &\sum\limits_{i=1}^{N} \int
 dE\,dl\,db\,~\frac{\delta(E-E_i)~\delta(l-l_i)~\delta(b -
   b_i)}{\mathcal{E} (E,l,b, t_{\rm start}, t_{\rm end})}\nonumber
 \\
&~\times\mathcal E(E,l,b,t_b,t_b+1500~\mathrm{s}) ~\sigma_e^{E_0}(E)~\sigma_b(E,v,l,b)\,.
\end{align*}

Delta-functions will be integrated and $E,l,b$ will
be replaced by $E_i,l_i,b_i$. Sigma function will require to sum only
the photons within ROI with energies larger than $E_0$. We finally get:
\begin{align*}
B_{E_0} = \sum\limits_{v=0}^{1} \sum\limits_{i=1}^{N}
&\frac{\mathcal E^v(E_i,l_i,b_i,t_b,t_b+1500~\rm{s})}{\mathcal
E(E_i,l_i,b_i,t_{\rm start},t_{\rm end})}\\
&\times \sigma_e^{E_0}(E_i)~\sigma_b(E_i,v,l_i,b_i)\,,\\ \nonumber
\end{align*}
where the sum formally goes through all LAT detected photons, but
effectively only  ones in the spatial ROI do contribute.

%%%%%%%%%%%%%%%%%%%%%%%%%%%%%%%%%%%%%%%%%%%%%%%%%%%%%%%%%%%%%%%%%%%%%%%%%%%%%%%%%%%%%%%
%%%%%%%%%%%%%%%%%%%%%%%%%%%%%%%%%%%%%%%%%%%%%%%%%%%%%%%%%%%%%%%%%%%%%%%%%%%%%%%%%%%%%%%%

\bibliographystyle{mn2e}
%\bibliography{GRB}

\end{document}